\documentclass[12pt]{article}
\usepackage{amsfonts}
\usepackage{psfig}
\newcommand{\beq}{\begin{equation}}
\newcommand{\eeq}{\end{equation}}
\newcommand{\beqn}{\begin{eqnarray}}
\newcommand{\eeqn}{\end{eqnarray}}
\newcommand{\bea}[1]{\beq\begin{array}{#1}}
\newcommand{\eea}{\end{array}\eeq}

\newcommand{\Tr}[1]{\;{1\over #1}\mathop{\rm Tr}}

\newcommand{\Pexp}{\mbox{P}\!\exp}

\newcommand{\ket}[1]{|\,#1\,\rangle}
\newcommand{\bra}[1]{\langle\,#1\,|}
\newcommand{\braket}[2]{\langle\,#1\,|\,#2\,\rangle}

\newcommand{\diff}{\partial}
\newcommand{\cC}{{\cal C}}

\newcommand{\NP}[3]{{\it Nucl. Phys. }{\bf #1} (#2) #3}
\newcommand{\NPPS}[3]{{\it Nucl. Phys. Proc. Suppl. }{\bf #1} (#2) #3}
\newcommand{\PL}[3]{{\it Phys. Lett. }{\bf #1} (#2) #3}
\newcommand{\PRL}[3]{{\it Phys. Rev. Lett. }{\bf #1} (#2) #3}
\newcommand{\PRep}[3]{{\it Phys. Rep. }{\bf #1} (#2) #3}
\newcommand{\PR}[3]{{\it Phys. Rev. }{\bf #1} (#2) #3}
\newcommand{\JL}[3]{{\it JETP Lett. }{\bf #1} (#2) #3}
\newcommand{\CMP}[3]{{\it Comm. Math. Phys. }{\bf #1} (#2) #3}

\newcommand{\IJMP}[3]{{\it Int. J. Mod. Phys. }{\bf #1} (#2) #3}

\newcommand{\PTP}[3]{{\it Prog. Theor. Phys. }{\bf #1} (#2) #3}
\newcommand{\PTPS}[3]{{\it Prog. Theor. Phys. Suppl. }{\bf #1} (#2) #3}
\newcommand{\JHEP}[3]{{\it JHEP }{\bf #1} (#2) #3}

\begin{document}
\date{}
\title{Gauge Invariant Monopoles\\
in SU(2) Gluodynamics.
\vskip-40mm
\rightline{\small ITEP-LAT-2002-03}
\rightline{\small MPI-PhT-2002-09}
\vskip 40mm
}
\author{F.V.~Gubarev$^{\rm a}$, V.I.~Zakharov$^{\rm b}$ \\
\\
$^{\rm a}$ {\small\it Institute of Theoretical and  Experimental Physics,}\\
{\small\it B.Cheremushkinskaya 25, Moscow, 117259, Russia}\\
$^{\rm b}$ {\small\it Max-Planck Institut f\"ur Physik,}\\
{\small\it F\"ohringer Ring 6, 80805 M\"unchen, Germany}
}

\maketitle
\thispagestyle{empty}
\setcounter{page}{0}
\begin{abstract}
\noindent
We introduce a gauge invariant topological definition of monopole
charge in pure $SU(2)$ gluodynamics. The non-trivial topology
is provided by hedgehog configurations of the non-Abelian field
strength tensor on the two-sphere surrounding the monopole. It is
shown that this definition can be formulated entirely
in terms of Wilson loops which makes the gauge invariance manifest.
Moreover, it counts correctly the monopole charge
in case of spontaneously broken gauge symmetry and of pure Abelian gauge fields.
\end{abstract}

\newpage
\subsection*{Introduction}
\noindent
The issue of magnetic monopoles in gauge models has a long history. 
The most famous result remains the Dirac quantization condition
\cite{Dirac}. Although the monopoles themselves appeared
to be purely mathematical constructions there were many 
exciting theoretical issues elucidated in the next  50 years or so
following the Dirac's paper,
for a review see~\cite{coleman}. The next breakthrough 
was the 
realization~\cite{HP} that magnetic monopoles naturally appear as 
classical configurations in the context of grand unified theories.

Since then the main development is the observation and accumulation of data
on the monopoles in lattice gauge theories (see Ref.~\cite{reviews} for a review).
The monopoles are intrinsically a $U(1)$ object
and in the non-Abelian case, with no spontaneous symmetry breaking,
the choice of a particular $U(1)$ turns out to be a matter
of gauge fixing. As a result, the
definition of the monopoles is not unique
and it is a separate issue which monopoles are relevant physically.

Despite of the gauge dependence of monopoles in non-Abelian models
the crucial point is that it is only the compactness of the gauge group that
makes the very existence of monopoles possible \cite{Pol}. Therefore the lattice regularization
is singled out since it explicitly preserves the global structure of
the gauge group. In particular, any non-Abelian
lattice action is a periodic functional which permits for this reason existence
of chromo-magnetic singular fluxes (Dirac strings)~\cite{strings,towards}.
In this paper we show that the monopoles defined as the end-points of
non-Abelian Dirac strings are $SU(2)$ gauge invariant.
It is important that the singularity of the non-Abelian flux does not necessarily
imply singular gauge potentials, contrary to the Abelian case.
Whether the potentials are singular or not depends now on a particular
gauge. We also formulate the monopole charge in terms of physical fluxes
(Wilson loops) alone thus demonstrating explicitly its gauge invariance.

To substantiate our definition of the magnetic charge 
we first consider a particular case of 
spontaneously broken gauge symmetry, namely, the Georgi-Glashow model and show that
at the level of classical field configurations our formulation is identical
to the well known definition of the 't~Hooft-Polyakov monopole charge. Then we
rederive it in terms of Wilson loops which allows us to check that in
the limit of pure Abelian gauge fields our construction is the same
as for the Abelian monopole. At the level of classical
field configurations it is also possible to estimate the self-energy of the
monopole, which turns out to be linearly divergent in the ultraviolet.

Thus, our definition of the monopole
charge if applied to classical field configurations
would select the singular Wu-Yang monopole~\cite{WuYa}. However, it is well known that
the Wu-Yang solution is in fact unstable~\cite{instability}. Physically
the instability arises because the  interaction
of spins with the magnetic field is so strong that the 
massless gluons fall onto the monopole
center. Thus, it is an open dynamical question, what kind of
the field configurations would be primarily identified as having
the magnetic charge. Most probably, one should rely on the numerical
simulations to answer this question. Thus, it is worth mentioning
that our, pure topological definition of the monopole charge
can be implemented in the quantum context~\cite{self} of the lattice gauge theories.

\subsection*{1. Monopole Charge in Georgi-Glashow Model.}
\noindent
Magnetic monopoles of finite energy were found \cite{HP} as classical solutions
to non-Abelian field theories with scalar matter fields. And in fact it is
a common belief that monopoles in QCD, if they exist, should resemble somehow the 
monopoles of the Georgi-Glashow model. 
However, this feeling is difficult to formalize 
since there are no scalar fields anymore.
To substantiate our definition of monopoles in gluodynamics it is convenient
to start from the Georgi-Glashow model where the monopole charge is a well defined quantity
which explicitly depends however upon the scalar fields profile.
We will reformulate the standard definition in terms of the gauge field alone. 
The latter construction then directly applies
in gluodynamics since it makes no reference to the scalar fields.
Of course, the equivalence of the both constructions holds only on the classical level.
As far as the quantum problem is concerned they produce {\it a priori} different results. 

Consider the Georgi-Glashow model with a scalar triplet $\Phi^a$:
\beq
L ~=~ \frac{1}{4g^2} (F^a_{\mu\nu})^2 ~+~ \frac{1}{2}(D_\mu \Phi^a)^2 ~+~ V(|\Phi|)\,,
\eeq
$$
F^a_{\mu\nu} ~=~ \diff_{[\mu} A^a_{\nu]} + \varepsilon^{abc} A^b_\mu A^c_\nu\,,
$$
$$
(D_\mu\Phi)^a ~=~ \diff_\mu\Phi^a ~+~ \varepsilon^{abc} A^b_\mu \Phi^c\,,
$$
where the potential $V(|\Phi|)$ is such that classically $\langle |\Phi| \rangle = v \ne 0$.
The 't~Hooft-Polyakov monopole solution corresponds to the ansatz:
\beq
\label{HP1}
A^a_i ~=~ \varepsilon^{aik} \frac{r^k}{r^2} A(r)\,, \qquad  \Phi^a = \frac{r^a}{r} v \Phi(r)\,.
\eeq
Moreover, let us consider sufficiently large vector- 
and Higgs- boson masses, (denoted $m_V$ and  $m_H$, respectively). Then
the following asymptotic behavior of the profile functions $A(r)$, $\Phi(r)$ is true almost everywhere:
\beq
\label{HP2}
\lim\limits_{r\to\infty} A(r) ~=~ \lim\limits_{r\to\infty} \Phi(r) = 1\,.
\eeq
Of course, Eqs.~(\ref{HP1},\ref{HP2}) are valid only for a single monopole.
In general, there are many objects like (\ref{HP1},\ref{HP2}) in the vacuum and the
question is, what is the net monopole charge contained in a
three-dimensional volume large compared to $1/m_V^3$, $1/m_H^3$?
The answer is that one has to take a closed smooth surface $S^2_{phys}$ containing the volume
and consider the Higgs field distribution on it. 
Then the net monopole charge can be calculated as:
\beqn
\label{HPcharge}
q = \frac{1}{8\pi} \int\limits_{S^2_{phys}} 
\hat{\vec{\Phi}} \cdot \diff \hat{\vec{\Phi}} \times\diff\hat{\vec{\Phi}} ~=~
\frac{1}{8\pi} \int\limits_{S^2_{phys}} 
\left\{
\diff\wedge \hat{\vec{\Phi}}\vec{A} ~+~ \hat{\vec{\Phi}} \cdot \diff \hat{\vec{\Phi}} \times\diff\hat{\vec{\Phi}}
\right\} ~=~ 
\\
\nonumber
\frac{1}{8\pi} \int\limits_{S^2_{phys}} 
\left\{
-\hat{\vec{\Phi}}\vec{F} ~+~ 
\hat{\vec{\Phi}} \cdot D\hat{\vec{\Phi}} \times D\hat{\vec{\Phi}}
\right\}\,,
\eeqn
where $\hat{\vec{\Phi}} = \vec{\Phi}/|\Phi|$. Thus the monopole trajectory is parameterized by
three equations $\Phi^a = 0$ which are quite convenient to use when the scalar fields are 
explicitly given. But suppose that Higgs field distribution is unknown. How can one detect
the monopole (\ref{HP1},\ref{HP2}) in this case?

The answer is straightforward provided that the radius of $S^2_{phys}$ is sufficiently large.
Indeed,  
the scalar field $\vec{\Phi}$ satisfying the classical
field equations is covariantly constant at large distances:
\beq
\label{HPconst}
D_i \Phi ~=~ 0 \qquad \left(\,rm_V, r m_H \gg 1\,\right)\,.
\eeq
Thus, the direction of the vector $\vec{\Phi}$ on $S^2_{phys}$ is the same as the direction
of the tangential to $S^2_{phys}$ components of $F^a_{\mu\nu}$:
\beq
\label{parallel}
\left. \vec{\Phi} \times \vec{F}_{\mu\nu} d^2\sigma^{\mu\nu}\right|_{S^2_{phys}} ~=~ 0\,,
\eeq
where $d^2\sigma^{\mu\nu}$ is an infinitesimal surface element. 
Therefore, assuming that $\vec{F}\,d^2\sigma \ne 0$
one can introduce the unit vector~\cite{strings,towards}
\beq
\label{firstN}
\vec{n} ~\sim~ \vec{F}d^2\sigma\,, \qquad \vec{n}\in S^2_{phys}
\eeq
and calculate the monopole charge $q$ by substituting $\hat{\vec{\Phi}}\to\vec{n}$
in Eq.~(\ref{HPcharge}):
\beq
\label{HPchargeNew}
q = \frac{1}{8\pi} \int\limits_{S^2_{phys}} 
\vec{n} \cdot \diff \vec{n} \times\diff\vec{n}\,.
\eeq
It is evident that in case of the classical 't~Hooft-Polyakov solution
Eqs.~(\ref{HPcharge}) and (\ref{HPchargeNew})
are exactly equivalent at least for sufficiently large
$S^2_{phys}$. What might be more surprising is that 
the equivalence of (\ref{HPcharge}) and (\ref{HPchargeNew})
holds true at all the distances, that is even inside the core of 
the 't~Hooft-Polyakov solution.
Indeed, the color magnetic field of the monopole (\ref{HP1},\ref{HP2}) is given by:
\beq
\label{junk}
B^a_k ~\sim~  \delta^{ak} A'/r ~+~ \frac{\hat{x}_a\hat{x}_k}{r^2}\,\left[ A - r A'\right]
\eeq
and therefore
\beq
\label{ncheck}
n^a ~\sim~ B^a_k ds^k ~=~ \hat{x}_a \, \frac{A}{r^2}\,,
\qquad
ds^k ~\sim~ \varepsilon^{kij} d^2\sigma_{ij}\,.
\eeq
It is clear that the monopole charge (\ref{HPchargeNew}) calculated on
the configuration (\ref{junk}) is exactly the same as the charge 
defined by Eq.~(\ref{HPcharge}).

We would like to emphasize that Eqs.~(\ref{firstN},\ref{HPchargeNew}) by themselves make no reference
neither to the Georgi-Glashow model nor to a particular classical solution.
They may also be considered in case of pure $SU(2)$ gluodynamics as well.
Therefore Eqs.~(\ref{firstN},\ref{HPchargeNew}) might serve
as a natural gauge invariant definition of the monopole charge in pure gauge theories. 

Let us make several remarks concerning the above monopole-charge definition:

{\it i)} The monopole charge is well defined as far as the unit vector $\hat{n}$
is defined,
that is as far as $\vec{F}d^2\sigma^{\mu\nu}$ is not vanishing.
However, even if  $\vec{F}d^2\sigma^{\mu\nu}=0$ for a particular $d^2\sigma^{\mu\nu}$
it does not mean that there is a magnetic monopole located at this point.
Generally speaking, it is enough to vary the surface. Indeed, vanishing of all
the components $F^a_{\mu\nu}$ constitutes too many equations to be satisfied on a world-line
in case of $D=4$. 

{\it ii)} One can verify that Eq.~(\ref{HPchargeNew}) is invariant
under small deformation of the surface $S^2_{phys}$ provided that the field (\ref{firstN})
is non singular. Therefore the Gauss law holds true for the magnetic charge (\ref{HPchargeNew}),
at least for smooth non vanishing $F^a_{\mu\nu}$.

{\it iii)} The definition (\ref{firstN},\ref{HPchargeNew}) makes no explicit reference to the non-Abelian
action density. In fact this is a common feature of all monopole constructions considered so far
in gluodynamics. What is specific, however, is that Eq.~(\ref{HPchargeNew}) is gauge invariant.
It is well known that the only gauge invariant objects in pure gluodynamics are the Wilson loops.
Therefore it should be possible to reformulate Eqs.~(\ref{firstN},\ref{HPchargeNew}) in terms
of Wilson loops alone. We consider this issue in the next section.

\subsection*{2. Monopole Charge in Terms of Wilson Loops.}
\noindent
In this section we reformulate the definition (\ref{firstN},\ref{HPchargeNew})
in terms of Wilson loops (see also Ref.~\cite{Berlin}).
In fact, our approach is close to quantum mechanical consideration of non adiabatic
Berry phase~\cite{berry} (see Ref.~\cite{berry-review} for a review).
Therefore we recall first the basics of the Berry phase construction and then
apply it to the $SU(2)$ pure gauge theory
(for related ideas see Ref.~\cite{OurBerry}).

\subsubsection*{2.1. Non Adiabatic Berry Phase.}
\noindent
Consider unitary evolution of a quantum mechanical system described by a state vector
$\ket{\psi(t)}$, where $\ket{\psi}$ is an element of the $N+1$-dimensional complex vector
space ${\mathbb C}^{N+1}$ with $N$ finite, for simplicity. In terms of the complex coordinates
$\{z_0,...,z_N\}$ the state vector is $\ket{\psi(t)}=\{z_0(t),...,z_N(t)\}$. Since the unitary
evolution preserves the norm, the normalization condition $\braket{\psi}{\psi}=1$ defines a
$2N+1$-dimensional sphere $S^{2N+1}\in{\mathbb C}^{N+1}$ on which the evolution takes place. 
However, the space of physical states is even narrower
since the states $\ket{\psi}$ and $\ket{\psi\,'}$ are physically equivalent if 
$\ket{\psi} = e^{i\alpha}\,\ket{\psi\,'}$. Therefore, the set of physically
non-equivalent states is the
$N$-dimensional projective space:
\beq
\label{CPN}
{\mathbb CP}^N ~=~ S^{2N+1}/\,U(1)\,.
\eeq
The quantum mechanical evolution is governed by the Schr\"odinger equation
\beq
\label{Schrodinger}
i \diff_t \ket{\psi(t)} ~=~ H(\,\lambda(t)\,) \, \ket{\psi(t)}\,,
\eeq
where Hamiltonian $H(\lambda(t))$ depends implicitly on time through the set of the parameters
$\lambda_i(t)$, $\lambda_i(0) = \lambda_i(T)$. We will be interested in the eigenstate
of the full evolution operator which returns to the same physical state  after evolving along
a closed path $C \in {\mathbb CP}^N$:
\beq
W(T)\, \ket{\psi(0)}~\equiv~ \ket{\psi(T)} ~=~ e^{i\varphi(T)}\,\ket{\psi(0)}\,,
\qquad
W(t)~=~ \mbox{T}\!\exp\{-i\int\limits^{t}_{0}H\}\,,
\eeq
Here $C$
is the image of a closed path $\cC = \{\lambda(t),0<t<T\}$ in the space of parameters $\lambda$
under the mapping $\ket{\psi(t)}:\,\cC\to C$.
In order to calculate the  total phase $\varphi(T)$ acquired by $\ket{\psi(0)}$ during the evolution,
we introduce a single-valued vector $\ket{\tilde{\psi}(t)}$
which differs from $\ket{\psi(t)}$ by a phase factor
\beq
\label{single}
\ket{\psi(t)} ~=~ e^{i\varphi(t)} \; \ket{\tilde{\psi}(t)}
\eeq
and satisfies the condition:
\beq
\label{single-valued}
\ket{\tilde{\psi}(T)} ~=~ \ket{\tilde{\psi}(0)}\,.
\eeq
Note that Eq.~(\ref{single-valued}) still does not define the $\ket{\tilde{\psi}}$ uniquely:
for a given $\ket{\tilde{\psi}}$,  vector $e^{i\theta}\;\ket{\tilde{\psi}}$ is also
single-valued provided that $\theta(T)-\theta(0)=2\pi n$.

From Eqs.~(\ref{Schrodinger},\ref{single}) it is straightforward to get the total phase:
\beq
\label{nonadiabatic-total}
\varphi(T)~=~ \delta ~+~ \gamma ~=~ - \int\limits_\cC \bra{\tilde{\psi}} H \ket{\tilde{\psi}}
~+~ i \int\limits_\cC \bra{\tilde{\psi}} d \ket{\tilde{\psi}}\,.
\eeq
The first term in Eq.~(\ref{nonadiabatic-total}) is known as the non adiabatic
dynamical phase. It depends explicitly  on detailed structure of the Hamiltonian. One can
show that in the adiabatic limit it reduces to $\delta=-\int_0^T E_n$
where $E_n$ is the corresponding energy level.
The second term is the non adiabatic Berry phase which depends only on
the form of the closed path $C\in{\mathbb CP}^N$ spanned by $\ket{\tilde{\psi}}$ during the
evolution.

Let us illustrate the above representation by the spin-$1/2$  evolution in an external
time-dependent magnetic field $\vec{B}(\lambda(t))$, $\lambda(T) = \lambda(0)$.
The corresponding Hamiltonian is $H = \frac{1}{2} B^a \sigma^a$ where
$\sigma^a$ are the Pauli matrices. A convenient basis is provided by the spin coherent
states $\ket{\vec{n}}$ which are in one-to-one correspondence with points on
two-dimensional sphere.
Note that the phase ambiguity for the states $\ket{\vec{n}}$ is fixed by the requirement that
every $\ket{\vec{n}}$ is obtainable via action of a specific $SU(2)$ element
on the highest weight state $\ket{1/2,1/2}$:
\beq
\braket{\theta,\phi}{\vec{n}}
~=~ \bra{\theta,\phi} \, \exp\{ i \frac{\theta}{2} \vec{m}\vec{\sigma}\}\,\ket{1/2,1/2}\,,
\qquad
\vec{m} = (\cos\phi\,,\sin\phi\,,0)\,.
\eeq
One can find the evolving family $\ket{\vec{n}(t)}$ from the defining equation,
\beq
e^{i\varphi(t)} \, \ket{\vec{n}(t)} ~=~ 
W(t)\,\ket{\vec{n}(0)} ~=~ 
\mbox{T}\!\exp\{-\frac{i}{2}\int\limits^{t}_{0}\vec{B}\vec{\sigma}\}\,
\ket{\vec{n}(0)}\,,
\eeq
where the initial vector $\ket{\vec{n}(0)}$ is an eigenstate of the full evolution operator
$W(T)\ket{\vec{n}(0)} = e^{i\varphi(T)} \ket{\vec{n}(0)}$. 
Let $\cC$ denote a closed path $\lambda(t)$, $0<t<T$ in space of the parameters $\lambda$.
Then the phase angle $\varphi(T)$ can be calculated as:
\beq
\label{angle1}
\varphi(T) ~=~ - \frac{1}{2}\oint\limits_\cC \bra{\vec{n}} \vec{B}\vec{\sigma} \ket{\vec{n}}
~+~ \frac{i}{2} \oint\limits_\cC \bra{\vec{n}} d \ket{\vec{n}}
~=~ 
- \frac{1}{2}\oint\limits_\cC \vec{B}\vec{n}
~+~ \frac{1}{4} \int\limits_{S_\cC} \vec{n}\cdot\diff\vec{n}\times\diff\vec{n}\,,
\eeq
where $S_\cC$ is a surface in the parameter space spanned on $\cC$. Note that
only $\varphi\,\mathrm{mod}\,2\pi$ is a well defined quantity.

\subsubsection*{2.2. Single Wilson Loop.}
\noindent
There is a particular class of quantum mechanical systems for which 
the space of state vectors carries unitary irreducible
re\-pre\-sen\-ta\-tion of a com\-pact semi\-simple Lie group $G$.
Namely, let the Hamiltonian $H$ be an element of the Lie algebra of $G$.
The time dependence is introduced, as usual, in terms of the parameters $\lambda_i(t)$.
Then the evolution operator for the Schr\"odinger equation (\ref{Schrodinger}),
\beq
\label{evolution-V}
V(t)~=~ \mathrm{T} \exp\{ -i \int_0^t H\}\,,
\eeq
is given by a path $[0;T]\to G$ in the group space and for any $t$ is an operator
in the representation space of $G$. Therefore, if at $t=0$ we start with an arbitrary state
$\ket{\psi(0)}$, then the state vector at the time $t$, $\ket{\psi(t)}=V(t)\ket{\psi(0)}$
is a generalized coherent state \cite{Perelomov}. 
Moreover, in case $G=SU(2)$ the generalized coherent states coincide in fact with the 
spin coherent states described above.

The evolution operator (\ref{evolution-V}) is of particular importance in gauge theories.
Indeed, consider a Wilson loop $W(T)$ in the fundamental representation
on the contour $\cC$ parameterized by coordinates $x_\mu(t)$, $t\in [0;T]$:
\beq
\label{Wilson-loop}
W(T) ~=~ \Pexp\{ i\int_0^T A(t) \; dt \}\,, \qquad
A(t)~=~ \frac{1}{2} A^a_\mu(x(t)) \;\dot{x}_\mu(t)\; \sigma^a \,,
\eeq
where  the dot denotes differentiation  with respect to $t$.  The P-exponent is
defined as a solution to the following first order differential equations:
\beq
(\; i\diff_t ~+~ A\;) \ket{\psi} ~=~0\,, \qquad
\ket{\psi(t)}~=~ W(t)\;\ket{\psi(0)}\,.
\eeq
Therefore the Wilson loop can be thought of as the evolution operator (\ref{evolution-V}) with  a
time-dependent Hamiltonian, $H=-A(x(t))$. Moreover, the space of parameters $\lambda$
is identified now with the physical space $x_\mu$.
The importance of the cyclic states and the corresponding
phase angles is also apparent since $\Tr{2} W(T) ~=~ \cos\varphi(T)$. 
Therefore one can directly apply the considerations of the previous section to the $SU(2)$
gluodynamics. Namely, any Wilson loop operator $W(t)$ gives rise to a family
of spin states $\ket{\vec{n}(t)}$
\beq
\label{family}
e^{i\varphi(t)} \, \ket{\vec{n}(t)} ~=~ W(t)\,\ket{\vec{n}(0)}\,,
\qquad
\vec{n}(T) ~=~ \vec{n}(0)\,,
\eeq
which map the contour $\cC$ in the physical space to a closed path $\cC_{color}$
on a unit two-dimensional sphere $S^2_{color}$. For the phase of the Wilson loop $\varphi$
one gets expression similar to Eq.~(\ref{angle1}):
\beq
\label{phase}
\varphi(T) ~=~ \frac{1}{2}\int\limits_\cC \vec{A}\vec{n}
~+~ \frac{1}{4} \int\limits_{S_\cC} \vec{n}\cdot\diff\vec{n}\times\diff\vec{n}\,,
\eeq
where $S_\cC$ is an arbitrary surface spanned on $\cC$. The second term in Eq.~(\ref{phase})
is nothing else but the oriented  solid angle corresponding to $\cC_{color}$ on $S^2_{color}$.

Of course, Eq.~(\ref{phase}) is by no means helpful in calculating the Wilson
loop itself since the construction of evolving spin states (\ref{family}) already requires the
knowledge of $W(t)$. Nevertheless, it might be useful in the context of the lattice 
gauge models where $W(t)$ can be evaluated by a direct multiplication of the link
matrices~\cite{self}.

Note that the phase angle $\varphi$ (\ref{phase}), as it stands, is {\it not}
a well defined quantity. The ambiguity in $\varphi$ stems from the freedom
to arbitrarily choose $S_\cC$ and from the gauge invariance of the Wilson loop.
Indeed, various choices of $S_\cC$ result in different values of $\varphi$ which differ
by $2\pi n$, $n\in Z$. On the other hand, a gauge transformation $g(t)$ applied along
$\cC$ changes the evolving family of spins $\ket{\vec{n}(t)} \to g(t)\ket{\vec{n}(t)}$. 
However, one can verify that in this case $\varphi(T)$ is changing by $2\pi n$ also
and therefore $\varphi\,\mathrm{mod}\,2\pi$ is gauge invariant and well defined.
From now on we will consider only $\varphi\,\mathrm{mod}\,2\pi$. Moreover, to simplify
notations the $\mathrm{mod}\,2\pi$ operation will not be indicated explicitly.

It is amusing to note that (\ref{phase}) looks like a semi-classical approximation
to the so called non-Abelian Stokes theorem~\cite{nast}:
\beq
\label{NAST}
\Tr{2} W(T) ~=~ \int\limits_{\vec{n}(T)=\vec{n}(0)}
D\vec{n}(t)\,\,
\exp\{\,\,\frac{i}{2}\int\limits_\cC \vec{A}\vec{n}+
\frac{i}{4} \int\limits_{S_\cC} \vec{n}\cdot\diff\vec{n}\times\diff\vec{n}\,\,\}\,.
\eeq
But in fact Eq.~(\ref{phase}) has little to do with any semi-classical evaluation of (\ref{NAST}).
Indeed, since equation (\ref{phase}) is exact it would imply the exactness of WKB approximation in the
theory (\ref{NAST}) which in turn is not the case (at least so to the best of our knowledge).
Thus the evolving family of spins (\ref{family}) does not correspond to any classical trajectory
in (\ref{NAST}).

To conclude this section we would like to mention 
that for any given Wilson loop there exist of course two eigenstates, $\ket{\pm\vec{n}}$
('spin up' and 'spin down') with corresponding phases $\pm\varphi(T)$.
For a single Wilson loop there is no distinction between $\ket{\pm\vec{n}}$ and
either of them might be taken as the initial state $\ket{\vec{n}(0)}$.
While in the continuum limit this sign ambiguity can be self consistently resolved,
on the lattice it becomes a severe problem~\cite{self}.

\subsubsection*{2.3. Monopole Charge in Terms of Wilson Loops.}
\noindent
Consider now an arbitrary closed smooth surface $S^2_{phys}$ in the physical space and cover it 
with a set of infinitesimal patches. The area of each patch is $\delta\sigma_x$ which in turn
is bounded by an infinitesimal contour $\delta\cC_x$. The phase angle evaluated on
$\delta\cC_x$ is:

\beq
\label{small-phase}
\delta\varphi_x ~=~ \frac{1}{4}\,\{\, \diff\wedge(\vec{n}\vec{A}) ~+~ 
\vec{n}\cdot\diff\vec{n}\times\diff\vec{n}\,\} \,\, \delta\sigma_x\,.
\eeq
It is tempting to integrate (\ref{small-phase}) over $S^2_{phys}$. But before doing this
one has to show that families of spin states $\ket{\vec{n}}$ constructed separately
on each $\delta\cC_x$ define a smooth vector field $\vec{n}\in S^2_{phys}$. 
In fact the smoothness of $\vec{n}\in S^2_{phys}$
immediately follows from the usual assumption of the gauge-fields continuity.
Indeed, the Wilson loop evaluated on an infinitesimal contour $\delta\cC_x$ is:
\beq
W(\delta\cC) ~=~ 1 ~+~ \frac{i}{2}\,(\sigma^a F^a_{\mu\nu})\, \delta\sigma_{\mu\nu}
~+~ O(\,(\delta\sigma)^2\,)
\eeq
and therefore the direction of $\vec{n}_x$ coincides (up to the sign, see below) with the
direction in the color space  of the tangential components of the field-strength tensor,
$\vec{F}\delta\sigma_x$.
Thus for smooth $\vec{F}\delta\sigma\in S^2_{phys}$ the vector field 
$\vec{n}_x$ is continuous as well. Moreover,
it is clear that $\vec{n}$ just constructed is the same as the unit vector field considered in
section 1, see Eqs.~(\ref{firstN},\ref{HPchargeNew}).

As a result we get the following expression for the monopole charge
contained in $S^2_{phys}$
which is written now in terms of infinitesimal Wilson loops alone:
\beq
\label{charge}
q ~=~ \frac{1}{2\pi} \oint\limits_{S^2_{phys}} \delta\varphi  ~=~
\frac{1}{8\pi} \oint\limits_{S^2_{phys}} 
\{\, \diff\wedge(\vec{n}\vec{A}) ~+~ \vec{n}\cdot\diff\vec{n}\times\diff\vec{n}\,\} \,\, \delta\sigma\,.
\eeq

Let us comment on the sign problem mentioned above. For every
infinitesimal Wilson loop $W(\delta\cC_x)$ there are two eigenstates $\ket{\pm\vec{n}_x}$
and either of them may be taken as $\vec{n}_x$.
But in fact for smooth $S^2_{phys}$ and continuous $\vec{F}\delta\sigma$ the sign
ambiguity is not important, because the continuity
requirement for the field $\vec{n} \in S^2_{phys}$
fixes completely the relative signs of $\vec{n}_x$ at neighboring $x$.
Therefore the only remaining freedom is to globally change the sign of all $\vec{n}_x$
which in turn is equivalent to changing the sign of $q$, see Eq.~(\ref{charge}).

Finally, let us note that for pure diagonal gauge fields the monopole charge (\ref{charge})
coincides with the well known Abelian monopole definition. Indeed, one can easily see that in this
case $n^a_x = \delta^{a3}$ everywhere on $S^2_{phys}$ and Eq.~(\ref{charge})
measures the Abelian magnetic flux of the monopole, as it should.

\subsection*{3. Monopoles and non-Abelian Dirac Strings.}
\noindent
Let us gauge rotate the field $\vec{n}\in S^2_{phys}$
to a fixed direction in the color space, e.g. $n^a =\delta^{a3}$.
Then the charge (\ref{charge}), at least on the classical level,
corresponds to the Abelian Dirac monopole embedded into $SU(2)$
along the diagonal subgroup. Moreover, it is accompanied
by the corresponding Dirac string which brings in the total flux
to the magnetic charge. Note that the standard Dirac string singularity appears
only in this particular gauge. Indeed, for the 't~Hooft-Polyakov monopole 
(\ref{HP1}) the potentials are regular
everywhere for $r\ne 0$. On the other hand, 
it seems reasonable to introduce such a generalization of the Dirac string
to the non-Abelian case that the string does not 
disappear while going from the unitary to hedgehog gauge.
Indeed, the magnetic monopoles are 
 intrinsically Abelian configurations
for which the magnetic flux is conserved.

One concludes, therefore, that it makes no much sense to discuss the Dirac strings
using the usual definition of the field strength tensor
\beq
\label{naiveF}
F_{\mu\nu}~=~ \diff_{[\mu} A_{\nu]} ~+~ [\,A_\mu \,,\,A_\nu]\,.
\eeq
Indeed, the chromo-magnetic field (\ref{naiveF})
is regular for continuous gauge potentials and no string-like singularity
can appear. On the other hand, there is another well known definition
of $F_{\mu\nu}$ which is introduced through consideration of matter field $\Phi_x$
in a given gauge-field background. Namely, the change in
$\Phi_x$ due to its transportation along an infinitesimal contour $\delta\cC_x$
is given by:
\beq
\label{F}
\Delta\,\Phi_x ~=~ i\,[\, F\delta\sigma_x\,,\,\Phi_x\,]\,,
\eeq
where $\delta\sigma_x$ is a surface spanned on $\delta\cC_x$ and 
$F\delta\sigma_x$ is the field strength tensor evaluated on $\delta\sigma_x$.
The advantage of Eq.~(\ref{F}) is that it allows to deal consistently  with large
chromo-magnetic fields, $|F\delta\sigma_x| \sim 1$, by considering
the evolution of $\Phi_x$ along $\delta\cC_x$. In fact, in the previous sections
we considered exactly the same Eq.~(\ref{F}) in the context of pure gauge theories.
Moreover, the states $\ket{\vec{n}_x}$ constructed above directly correspond to the
field $\Phi_x$ and the phase angle (\ref{small-phase}) coincides with
the magnitude of the color vector $\vec{F}\delta\sigma_x$.

The latter observation paves the way for a natural definition of the Dirac strings
in non-Abelian gauge models. Indeed, let us remind the reader that we are always considering
{\it not} the full phase of the Wilson loops $\varphi$, but $\varphi\,\mbox{mod}\,2\pi$
instead. One can readily see that the appearance of a non-zero magnetic charge in
Eq.~(\ref{charge}) is entirely due to this $\,\mbox{mod}\,2\pi$ operation\footnote{
Of course, one can glue $S^2_{phys}$ from several patches and follow the Wu-Yang
construction\cite{WuYang} to avoid singularities in $\delta\varphi_x$.
But this is essentially the same approach, since 
$\,\mbox{mod}\,2\pi$ operation would inevitably reappear when patches are glued together.
}. This means in particular, that once $q \ne 0$ there must be at least one point
$x_0$ on $S^2_{phys}$ for which
\beq
\label{string}
\lim\limits_{\delta\sigma_{x_0} \to 0} \delta\varphi_{x_0} ~=~ \pm 2\pi\,.
\eeq
In fact, the singular point $x_0$ is the location of the non-Abelian Dirac string
since it corresponds to a singular flux 
\beq
\label{flux}
F^a_{\mu\nu} F^a_{\mu\nu} ~\sim~ 4\pi^2 / a^2 \qquad
\mbox{no summation over $\mu$, $\nu$}
\eeq
piercing $S^2_{phys}$. Moreover, Eqs.~(\ref{string},\ref{flux}) reveal
a distinguished role of the lattice regularization~\cite{strings,towards}.
Indeed, on the lattice the action is intrinsically periodic:
\beq
\label{lataction}
S_{lat} ~\sim~ \frac{1}{g^2} \, \sum\limits_p \Tr{2} W_p ~=~
\frac{1}{g^2} \, \sum\limits_{\delta\sigma_x} \cos\delta\varphi_x\,,
\eeq
where the sum is taken over all elementary (infinitesimal) two-dimensional cells.
Therefore the Dirac strings (\ref{string},\ref{flux}) are automatically invisible
within the lattice formulation, in contrast with the conventional continuum gluodynamics
which completely suppresses by the action factor the fluctuations (\ref{flux}). 
Therefore the consideration
of monopoles in usual gluodynamics is in fact inconsistent unless extra rules
to deal with the string singularities (\ref{flux}) are introduced
which would bring it in line with the lattice formulation.

Similar to the Abelian case, one can show that the Dirac string is
completely unphysical object because its location can be arbitrarily shifted
by an appropriate gauge transformation. However, in the non-Abelian models the
string (\ref{string}) is even more of a phantom since it does not necessarily imply
singularities of gauge potentials. This can be explicitly demonstrated
for 't~Hooft-Polyakov monopole solution (\ref{HP1}) in the hedgehog and unitary gauges.
The only important point is that for any non-zero magnetic charge (\ref{charge})
there is at least one Dirac string which brings in the total chromo-magnetic
flux to the monopole.

\subsection*{4. Monopole Self-Energy.}
\noindent
The arguments above show that it makes no much sense to ask what
is the self-energy of the magnetic monopoles (\ref{charge}) within the conventional gluodynamics.
Indeed, the main contribution would come in this case from the self-energy of
the Dirac string which is quadratically divergent in the
ultraviolet. But once the string energy is subtracted (either by
hands or via the lattice regularization), the self-energy of the monopole (\ref{charge})
becomes well defined and may be estimated by rewriting Eq.~(\ref{charge}) in an
explicitly gauge invariant form:
\beq
\label{charge2}
q ~=~ \frac{1}{8\pi} \oint\limits_{S^2_{phys}} 
\{\, - \vec{n} \vec{F} ~+~ \vec{n}\cdot D\vec{n} \times D\vec{n}\,\} \,\, \delta\sigma\,,
\eeq
where $\vec{F}$ is the full non-Abelian field strength tensor and $D$
is the covariant derivative. By construction, the vector field $\vec{n}_x$ is
everywhere parallel to $\vec{F}\delta\sigma_x$  and therefore 
\beq
\label{zero}
[ D\vec{n} , D\vec{n} ] ~=~ 0\,, \qquad \vec{n}\cdot D\vec{n} \times D\vec{n} ~=~ 0
\eeq
is true almost everywhere
(in fact, Eq.~(\ref{zero}) is violated at the position of the Dirac string (\ref{string}) where
$\vec{n}\cdot D\vec{n} \times D\vec{n} \sim 2\pi/\delta\sigma$). 
If we assume the rotational symmetry of the monopole (\ref{charge2}), then
\beq
\left(\, F^a_{\alpha\beta}\,\right)^2 
~\sim~ \left[\,\, \frac{2 q}{R^2}\,\,\right]^2
\qquad \mbox{no summation over $\alpha$, $\beta$}\,,
\eeq
where $R$ is the radius of $S^2_{phys}$ and $\alpha$, $\beta$ denote the tangential components
of $\vec{F}_{\mu\nu}$. The estimation of the monopole self-energy is then:
\beq
\label{self-real}
E_{self} ~\ge~  \frac{(2q)^2}{g^2}\cdot \frac{1}{a}\,,
\eeq
which might also be obtained in a slightly different way. Namely, since Eq.~(\ref{charge})
is gauge invariant and the contribution of the first term on the r.h.s. vanishes
it is possible to find such a gauge that
\beq
\label{g-const}
\gamma \equiv \vec{n}\cdot\diff\vec{n}\times\diff\vec{n} = \frac{2q}{R^2}
\eeq
for any two-dimensional sphere $S^2(R)$ of radius $R$ centered at the origin. Thus:
\beq
E_{self} \ge \int \frac{R^2}{4g^2} dR \int\limits_{S^2(R)} (F^a_{\alpha\beta})^2
\ge\int \frac{R^2}{4g^2} dR \int\limits_{S^2(R)} [\gamma^2+ 2\gamma\diff\wedge\vec{n}\vec{A}]\,.
\eeq
The latter integral is zero due to Eq.~(\ref{g-const}) and the estimate (\ref{self-real}) follows.

In view of the result (\ref{self-real}) it is natural to ask whether
the field configurations with a non-zero charge (\ref{charge})
are at all relevant dynamically. Indeed, the $g^{-2}$ factor indicates that we are dealing with a
classical solution which is, however, infinitely heavy as one would expect on the dimensional
grounds. In fact, it is straightforward to identify the 
corresponding classical field configuration
and it  is nothing else but the Wu-Yang monopole:
\beq
\label{WY} A^a_i ~=~ \varepsilon^{aik} \frac{r^k}{r^2}\,.
\eeq
Moreover, the quantum effects would result in the running of the coupling, $g^2\to g^2(a)$, 
where $a$ is the ultraviolet cut off. Such monopoles could not
condense, as it follows from the standard
action-entropy balance \cite{Pol}. Another setback for the semi-classical approximation to the monopoles
(\ref{charge}) is that  the solution (\ref{WY}) is known to be
unstable~\cite{instability} due to the strong magnetic interaction with massless
gluons. Thus, the field configurations with non-zero
charge (\ref{charge}) are to be thought of as quantum objects
and their anatomy is poorly understood at present (see also Ref.~\cite{anatomy}).
It is amusing to note that there exist strong indication~\cite{vortex}
that the monopoles on the lattice are {\it not} rotationally invariant
which was our starting assumption in the classical approximation above.

To summarize, there are no reasons to expect that classical configurations (\ref{WY})
survive in the real vacuum. This is clear already from the instability of the solution (\ref{WY}).
However, the topological definition (\ref{charge}) itself is not related to the
classical approximation. Eq.~(\ref{charge}) is sensitive only to the topological
properties and not to the concrete fields distribution. 
Analytically, it does not seem possible to predict which field configurations would
dominate the nontrivial $q$.
However, the monopole charge definition (\ref{charge}) can be used in the lattice
gauge models~\cite{self} and the monopole dynamics can be investigated numerically.

\subsection*{Conclusions}
\noindent
We have shown that in pure SU(2) gluodynamics it is possible to define
a gauge invariant monopole charge despite of the absence of physical
scalar fields. The non-trivial topology is provided by the non-Abelian
field strength tensor. Namely, the non zero monopole charge in a three
dimensional volume is due to the hedgehog configuration of $F_{\mu\nu}$
on the boundary of that volume. Loosely speaking the physical
Higgs field in our construction is replaced by the gauge fields tensor
itself.

We have shown that the topological definition reproduces the standard
monopole charge formulation in the Georgi-Glashow model and in the limit of pure
Abelian gauge fields. We were also able to represent the monopole charge
entirely in terms of Wilson loop thus demonstrating explicitly
its gauge invariance. Moreover, there is a natural analog of the Dirac strings
in our construction which are defined as a quantized singular non-Abelian
flux piercing infinitesimal surface element.
This allows to alternatively define the monopoles as end-points of
the non-Abelian Dirac string. While the string itself is not physical and
its position is not meaningful, the location of the string's
end-points is well defined and gauge invariant. Therefore in the
language of the non-Abelian Dirac strings our definition is not much different from
the corresponding Abelian construction. The only important
difference is that in the non-Abelian case the string singularity
does not imply singular gauge potentials.
Whether the potentials are singular, depends on choice of gauge.

We have also estimated at the classical level the self-energy of the monopole.
As could be expected, it is linearly divergent in the ultraviolet, the same as in pure
Abelian case. However, the corresponding classical solution (Wu-Yang monopole)
is much different from its Abelian akin since it is unstable. 
The instability reflects the non-Abelian nature of the theory and physically
corresponds to the gluons falling onto the center of monopole.
The monopole instability cannot be treated self-consistently
because of the strong back-reaction on the solution itself.
Thus we are driven in the direction of numerical lattice simulations
to investigate the properties of the configurations with a non-trivial
charge~(\ref{charge})~\cite{self}.

\subsection*{Acknowledgments.}
\noindent
We acknowledge thankfully fruitful discussions with  V.~Belavin,
M.N.~Chernodub,R.~Hofmann, M.I.~Polikarpov and
L.~Stodolsky. The work of F.V.G. was partially supported by grants
RFFI-02-02-17308, RFFI-0015-96786, INTAS-00-0011 and CRDF award RP1-2364-MO-02.
The work of V.I.Z. was partially supported by grant INTAS-00-0011 and DFG program
{\it 'Hadrons and the Lattice QCD'}.


\end{document}